\documentclass[a4paper,12pt]{amsart}

\usepackage{amssymb,amsmath,amsthm,amscd} 

\usepackage{mathrsfs}
\usepackage{tabularx}
\usepackage{enumerate} 
\usepackage{graphicx}

\def \R {\mathbb{R}}

\def \xmax {x_{max}}

\begin{document} 

\newtheorem{theorem}{Theorem}%[section]
\newtheorem{proposition}{Proposition}%[section]
\newtheorem{definition}{Definition}%[section]
\newtheorem{lemma}{Lemma}%[section]
\newtheorem{remark}{Remark}%[section]

\numberwithin{equation}{section}

\def\decdot{\mathalpha{\mbox{$\cdot$}}}

\title{A nonlocal model of epidemic network with nonlimited transmission: Global existence and uniqueness }
\author{E.Logak, I.Passat}
\begin{abstract}
Following \cite{ipel1}, we consider a nonlinear SIS-type nonlocal system describing the spread of epidemics on networks, assuming nonlimited transmission,  
We  prove local existence of a unique solution for any diffusion coefficients and global existence in the case of equal diffusion coefficients. 
Next we study the asymptotic behaviour of the solution and show that the disease-free equilibrium (DFE) is linearly and globally asymptotically stable when the total mean population is small. Finally, we prove that the solution of  the system converge to the $DFE$.
 \end{abstract} 

\address {Universit\'e de Cergy-Pontoise, Departement de Math\'ematiques et Laboratoire AGM (UMR CNRS 8088),
2 avenue A. Chauvin, 95302 Cergy--Pontoise Cedex, France}
\email{elisabeth.logak@u-cergy.fr, isabelle.passat@u-cergy.fr}

\maketitle

\setcounter{equation}{0}

\setcounter{equation}{0}

\section{Introduction}\label{S1}
In this paper, we consider the model proposed in \cite{ipel1} for the spread of epidemics on heterogeneous networks. It is
 formally derived from a discrete model proposed in \cite{sal08}, \cite{sal09} by considering the degree as a continuous variable. 
Precisely, each node of the network corresponds to a patch of the metapopulation and is characterised by its degree $x>0$. We assume that the network is described by a continuous labelling of the nodes and is characterized by a degree distribution with density  $p(x)$. We assume that $p$ has compact support $J= [0,x_{max}]$ with $x_{max} >0$ and satisfies the assumptions
\begin{equation}
\forall x \in J,\,\,\,\,p(x)  \geq 0,\,\,p(0)=0 \mbox{ and } \int_0^{\xmax} p(x) \, dx = 1 \label{proba} \\
\end{equation}
The condition $p(0)=0$ corresponds to a connected network, with no isolated patch.

\noindent
 We restrict to the case of uncorrelated network.
\noindent
For any function $\phi:J \rightarrow \R$ with $\phi \in L^1(J, p(x) dx)$, we denote by $\langle \phi \rangle \in \R$  its mean value on the network defined by
\begin{equation}\label{meanfi}
\langle \phi \rangle= \int_0^{\xmax}  \phi(x) \, p(x) dx .
\end{equation}
We  assume the following property
$$  \phi \geq 0  \mbox{ on } J, \,\,\, \langle \phi \rangle=0 \,\,\, \Rightarrow \,\,\, \phi =0  \mbox{ on } J.$$
We also assume that the first-order moment is finite and denote  by $m>0$ its value defined as
\begin{equation}\label{fmean}
 m= \langle x \rangle =\int_0^{x_{max}}  x \, p(x) \, dx.
\end{equation}
Let us denote by $S(x,t)$ (resp. $I(x,t)$)  the density of susceptible (resp. infected) individuals of degree $ x \in J$ at time $t \geq 0$. In our model, their evolution in time is given by the solution of the following  nonlinear and nonlocal system,
$$(C)
\left\{
\begin{array}{lcll}
\vspace{0.7mm}
\displaystyle{\frac{\partial S}{\partial t}}&=&I(\mu-\beta_0 S)-D_S(S-\displaystyle{\frac{x}{m} s(t) } ),& x \in J,\,  t>0\\
\vspace{0.5mm}
\displaystyle{\frac{\partial I}{\partial t}}&=&I(-\mu+\beta_0 S)- D_I(I - \displaystyle{\frac{x}{m} i(t) }),&  x \in J,\,  t>0\\
S(x,0)&=&S_0(x),&  x \in J\\
I(x,0)&=&I_0(x),&  x \in J
\end{array}
\right.\
$$
with $s(t)$ and $i(t)$ given by
\begin{equation}
\forall t>0,  \,\,\,  \,\,\,  s(t)=\langle S(.,t) \rangle, \,\,\,   \,\,\,  i(t)=  \langle I(.,t) \rangle . \label{si} 
\end{equation}
Here, $D_S, D_I>0$ are the diffusion coefficients and the constant $\mu>0$ is the recovery rate of the epidemics. 
We also define the total density of individuals of degree $x\in J $ at time $t\geq 0$ 
by 
\begin{equation}\label{defN}
N(x,t) = S(x,t) + I(x,t).
\end{equation}
In this paper, we deal with  the case of nonlimited transmission and always assume that the transmission rate is a constant $\beta_0>0$.
The companion paper \cite{ipel1} is devoted to the analysis of the model under the assumption of limited transmission, where 
$$\beta=  \frac{\beta _0}{N(x,t)}.$$
We'll see later that this assumption critically affects the results about equilibria.

% In the sequel, the mean values of the functions $S(.,t))$, $I(.,t))$, $N(.,t)$ at time $t\geq 0$ are defined respectively as follows.
%\begin{equation}
%n(t)=<N(.,t)>= \int_0^{+ \infty}  N(x,t) p(x) dx = s(t) + i(t) .\label{n}
%\end{equation}

\noindent
The system $(C)$ is formally derived from the discrete SIS metapopulation model proposed in \cite{sal08,sal09,sal10}.  We refer the reader to {\cite{ipel1} for a precise presentation of the discrete model.

\noindent
Many recent papers have investigated dynamic processes on complex networks using the tools and methods of 
statistical physics (see \cite{new03} for a survey). In particular, epidemic networks have recently attracted a lot of attention both from modelling and practical and from theoretical viewpoint (see \cite{pv01} and \cite{keeling05} for an introduction and \cite{ps15} for a recent review).

This paper and its companion paper \cite{ipel1} originated from a series of papers by R.Pastor-Satorras, A.Vespignani, V. Collizza (\cite{cpv07}, \cite{cv07},
\cite{cv08}) which focuses on the propagation of epidemics in  ''metapopulation'', a concept introduced to describe a population in which individuals  are spatially distributed in patches forming subpopulations. Hence the metapopulation is naturally mapped onto a network, where nodes correspond to patches of subpopulations that migrate along the edges. The transmission of the epidemics takes place in each patch, whose population is composed of susceptible and infected individuals. Following these works, J. Saldana proposed
in \cite{sal08} and \cite{sal09}  a discrete model for the propagation of epidemics
in a heterogeneous network, with the additional feature that reaction and diffusion processes take place simultaneously, allowing him to derive a time-continuous model. We propose to
consider a continuous version, where both time and degree take values in $\R_+$. 
We claim that the resulting model is retaining the main features of the discrete ones, while simplifying the analysis and improving some of the results. Namely, we prove rigorously global existence and uniqueness results for system $(C)$. Besides, we establish in \cite{h2equilibrium} the existence of a threshold for the existence of an endemic equilibrium and its stability, which depends on the whole degree distribution $p$, and not just on its maximum value as was obtained on the discrete model.

\noindent
 The plan of the paper is the following. In section 2,  we show that the system $(C)$ is well-posed  for any couple of strictly positive diffusion coefficients $(D_I, D_S)$ and with nonnegative smooth initial data.  In the case of equal diffusion coefficients, we prove that the solution is global in time. The result is obtained using fixed point methods and apriori estimates on the solutions. In section 3,  we investigate the linear stability of the disease-free  equilibrium (DFE) and prove that for small enough $n_0$, it is also globally asymptotically stable. Next in section 4, we prove that if $n_0$ small enough, the functions $(S,I)$ converge to the DFE.

\section{Solutions of system $(C)$}
\[  \]
In order to prove the existence of solutions to $(C)$, let $T>0$ be a fixed time and 
associate to a given couple of continuous nonnegative functions $(s,i)$ defined on $[0,T]$ 
the following evolution system 
$$(\widehat C)
\left\{
\begin{array}{lcll}
\vspace{0.7mm}
\displaystyle{\frac{\partial \widehat S}{\partial t}}&=&\widehat I(\mu-\beta_0 \widehat S)-D_S(\widehat S - \displaystyle{\frac{x}{m}}s(t))& x \in J, \, t \in [0,T]\\
\vspace{0.5mm}
\displaystyle{\frac{\partial \widehat I}{\partial t}}&=&\widehat I(-\mu+\beta_0 \widehat S)-D_I(\widehat I-\displaystyle{\frac{x}{m}}i(t))& x \in J, \,  t \in [0,T]\\
\widehat S(x,0)&=&S_0(x),& x\in J \\
\widehat I(x,0)&=&I_0(x),& x \in J
\end{array}
\right.
$$
and denote by ($\widehat S,\widehat I )$ its solution on $ J_T = J \times [0,T]$.
We also define 
$$ \forall (x,t) \in J_T, \,\,\,  \widehat N(x,t) = \widehat S(x,t) + \widehat I(x,t)$$ 
and for all $t\in [0,T]$ we define the functions $(\widehat{s},\widehat{i})$ by
\begin{equation}\label{sihat}
\forall t\in [0,T],\,\,\, \widehat{s}(t)= \langle   \widehat{S}(.,t) \rangle, \,\,\, \widehat{i}(t)= \langle   \widehat{I}(.,t) \rangle 
\end{equation}
and  accordingly the function $\widehat n$ by
$$\forall t\in [0,T],\,\,\,\widehat n(t)= \langle   \widehat{N}(.,t) \rangle= \widehat s(t) + \widehat  i(t).$$

\noindent
Note that $(\widehat{S},\widehat{I})$ is a solution of $(C)$ if and only if $\widehat s =s$ and $\widehat i=i$ so that the existence of a solution to $(C)$ follows from the existence of a fixed point to the operator $(s,i)\rightarrow  ( \widehat s,\widehat i)$.\\
We multiply each equation by $p(x)$ and integrate on $[0, \xmax]$ to obtain that $(\widehat s,\widehat i)$ satisfies
$$(\widehat c)
\left\{
\begin{array}{lcl}
\widehat{s}'&=&\mu \widehat{i} -\beta_0\int_0^{\xmax} (\widehat{S}\widehat{I})(x,t) \, p(x)\, dx- D_S\, \widehat{s}+D_S\, s(t)\\
\widehat{i}'&=&-\mu \widehat{i}+\beta_0 \int_0^{\xmax}(\widehat{S}\widehat{I})(x,t) \, p(x)\, dx-D_I\, \widehat{i}+D_I \ i(t)\\
\widehat s(0)&=&s_0\\
\widehat i(0)&=&i_0. 
\end{array}
\right.
$$

\subsection{A priori estimates on ($\widehat{S},\widehat{I}$)}\label{2.4.2}
We make the following assumptions on the initial data.
\begin{enumerate}\label{id}
\item[] {$(ID1)$} The initial data $S_0: J \rightarrow  \R_+$ and $I_0: J \rightarrow  \R_+$ are smooth, nonnegative functions on $J$ satisfying $S_0(0)=0$ and $I_0(0)=0$.
\item[] {$(ID2)$} We denote by $s_0= \langle S_0 \rangle $ (resp. $i_0= \langle I_0 \rangle$) the initial average value of $S$ (resp.  $I$) and by
$n_0=s_0+i_0$ the initial total average population.We assume that $0< s_0, i_0 < \infty$.
\end{enumerate}

We  assume that the initial data satisfy $(ID1)$ and $(ID2)$.
\begin{lemma}  \label{signeichapeau}
Let $T>0$ be given and consider 2  smooth, nonnegative functions $(s,i)$ defined on $[0,T]$.\\
Then system $(\widehat C)$ admits a unique smooth solution $(\widehat S,\widehat I)$ on $J_T$.
The following properties are satisfied.
\begin{itemize}
\item[]{(i)} $\forall (x,t) \in J_T$, $\widehat I(x,t) \geq 0 $
\item[]{(ii)} $\forall (x,t) \in J_T$, $\widehat S(x,t) \geq 0$
\item[]{(iii)} $\forall t \in [0,T]$,  $\widehat I(0,t)= \widehat S(0,t) = 0$
\end{itemize}
\end{lemma} 
\noindent
{\bf Proof of Lemma \ref{signeichapeau}}\\
The local existence and uniqueness of a solution ($\widehat{S},\widehat{I}$) of $(\widehat C)$ follows from classical results on system of odes.
To establish $(i)$, note that $\widehat I $ satifies \\ 
\begin{equation}\label{II} 
\frac{\partial \widehat I}{\partial t}= \widehat I [-(\mu + D_I)+\beta_0 \widehat S]+ D_I\frac{x}{m}i(t)
\end{equation}
We define the functions $\widehat f$ and $\widehat F$ by 
$$ \forall (x,t) \in J_T, \,\,\, \widehat f(x,t)=-(\mu+D_I) +\beta_0\widehat S(x,t) \mbox{ and }\widehat F(x,t)=\int_0^t \widehat f(x,s)ds .$$
 Thus for
 all $(x,t) \in J_T$,
$$\frac{\partial (e^{-\widehat F}\widehat I)}{\partial t}=e^{-\widehat F}[ -\widehat f \widehat I+ \frac{\partial \widehat I}{\partial t}]=e^{-\widehat F}D_I\frac{x}{m}i(t)\geq 0 .$$
Since $I_0(0)\geq 0 $, we deduce that the property $(i)$ holds.\\
To establish $(ii)$, note that the function $\widehat S$ satisfies  
\begin{equation}\label{ode S}
\frac{\partial \widehat S}{\partial t} = \widehat S[- \beta_0 \widehat I- D_S] +\mu \widehat I +D_S\frac{x}{m}s(t).\\
\end{equation}
Since $\widehat I \geq 0$ on $J_T $  by $(i)$, the same method as in the proof of $(i)$ allows us to conclude that $(ii)$ holds.\\
Note that $\alpha (t)= \widehat I(0,t)$ satifies  (\ref{II}) with $x=0$ which reads $\alpha' =f(0,t)\alpha $.
Since $\alpha (0)=0$ by $(ID1)$, it follows that $\forall t \in [0,T]$, $\alpha (t)=0.$\\
We prove similarly that for all $t \in [0,T]$, $\widehat S(0,t)=0.$\\
This completes the proof of Lemma \ref{signeichapeau}.

\subsection{Existence of solutions to $(C)$ in the case where $D_I=D_S$}\label{2.4.3}
Let us first recall a standard version of the Gronwall's Lemma that is used in the sequel.
\begin{theorem}[Gronwall's lemma]\label{LG}
Let us consider 2 $C^1$ functions $A: J \times \R _+  \rightarrow \R _+$ and $C: J  \rightarrow \R _+$.
Let us consider a $C^1$ function  $f: J  \times \R _+ \rightarrow \R$ such that  $f(x,0)=0$ for all $x \in J$ and such that
 
\begin{equation}\label{f'}
\forall t \in \R _+, \,\,\, \forall x \in J,\,\,\, \vert \frac{\partial f} {\partial t}(x,t)\vert \leq A(x,t) +C(x)\vert f(x,t)\vert
\end{equation}
Then, 
$$ \forall t \in \R _+, \,\,\, \forall x \in J,\,\, \vert f(x,t)\vert \leq  e^{C(x)t}\int_0^t e^{-C(x)u}A(x,u)du  .$$
\end{theorem}
 
We assume here that $D_I=D_S$ which allows us to obtain global existence using a simpler proof, namely a fixed point theorem for the operator ${i \mapsto \hat i}$ .
\begin{theorem}  \label{existence} 
For any $T>0$,
 there exists a unique solution to $(C)$ on $J_T $.
\end{theorem}
\noindent
{\bf Proof}\\
\noindent
We consider  the set
$$E =  \lbrace i \in C^0([0,T]) ,\,\,\forall t >0, \,\,\, 0\leq i(t)\leq n_0 \rbrace,$$
where $C^0([0,T])$ is equipped with the  norm $\Vert f \Vert =\displaystyle{\sup_{t\in [0,T]}e^{-\lambda t} \vert f(t)\vert }$,
for a suitable $\lambda >0$ to be defined later.
To any $i \in E$, we associate the function $s$ defined by
\begin{equation}
\forall t \in [0,T], \,\,\, s(t)+i(t)=n_0
\label{somme}
\end{equation}
so that $s \geq 0$ on $[0,T]$. 
Let us define $$\phi:\,\ E \longrightarrow C^0([0,T]) $$ 
                  $$       i          \longmapsto    \widehat i,$$ 
where $\widehat i$ is defined in (\ref{sihat}).
We prove below that  $ \phi (E)\subset E$ and that there exists $k \in (0,1)$ such that  
$\phi $ is a $k$-contraction in $E$. \\
The proof is decomposed in three steps.\\
{\bf a) For all $i \in E$, $\widehat i\, \in \ E$}

\noindent
This property relies on the following property.
\begin{lemma}\label{nochapeau}
$\forall t >0$, $\widehat{n}(t)=n_0$
\end{lemma}
{\bf Proof}
Let us add the two lines of system $(\widehat c)$. Using that $s(t)+i(t)=n_0$, 
we obtain that for all $t \in [0,T]$, 
$$ \widehat{n}'(t)= \widehat{s}'(t) + \widehat{i}'(t)=-D_I(\widehat{n}(t)-n_0).$$
Since  $\widehat{n}(0)= n_0$, it follows that for all $t \in [0,T]$,
 $\widehat{n}(t)=n_0$.
 
 \noindent
In Lemma \ref{signeichapeau} (i), we proved that $\widehat I\geq 0$ and $\widehat S\geq 0$ on $J_T$.
Thus $\widehat i(t)\geq 0$ and $\widehat s(t)\geq 0$  for all $t \in [0,T]$. Since $\widehat n(t) =\widehat s(t) + \widehat i(t)  = n_0$ by Lemma  \ref{nochapeau}, it follows that $ \forall t \in [0,T]$, $0 \leq \widehat i(t) \leq \widehat n(t)=n_0$, which proves that $\widehat i\, \in \ E$ and establishes a).

{\bf b) expression of $ \widehat N $ }\\

\begin{proposition} \label{Nexpression}
The function $\widehat N$  only depends on the initial data and is given by
$$ \forall  (x,t) \in J_T, \,\,\, \widehat N(x,t)=K(x)e^{-D_It }+n_0\frac{x}{m} \mbox { with }K(x)=N(x,0)-n_0\frac{x}{m}.$$
In particular, we have that
\begin{equation}
 \forall  (x,t) \in J_T,\,\,\, 0\leq \widehat N(x,t)\leq N(x,0) +n_0\frac{x}{m} \label{majhat}
\end{equation}
\end{proposition}
\noindent
{\bf Proof}\\
When we add the 2 lines of $(\widehat C)$, $ \widehat N$ satifies \\
$$\frac{\partial \widehat N}{\partial t}=\frac{\partial \widehat S}{\partial t}  +\frac{\partial \widehat I}{\partial t} 
  =-D_I(\widehat N-n_0 \frac{x}{m}) $$
  Thus  for all $x\geq 0$ and all $t\geq 0$, $$ \widehat N(x,t)= K(x) e^{-D_It} +n_0 \frac{x}{m}$$
with  $\widehat N(x,0)= K(x) +n_0 \frac{x}{m}$, which yields the above expression of  $\widehat N$.
Hence
$$ \widehat N(x,t)=N(x,0) e^{-D_I t }+n_0\frac{x}{m} ( 1 - e^{-D_I t }),$$
which implies (\ref{majhat}).

\noindent
{\bf c) $\phi$ is a  contraction in $E$. }
Let us consider a function $i_1\in E$ (resp. $i_2 \in E$) and denote by $(\widehat S_1, \widehat I_1)$ (resp. $(\widehat S_2, \widehat I_2)$) the corresponding solution of $(\widehat{C})$.\\
First we prove that the function
$y=\widehat I_2- \widehat I_1$ satisfies
$$ \forall (x,t) \in J_T, \,\,\, \vert y(x,t)\vert \leq D_I\frac{x}{m} \Vert i_1-i_2 \Vert \frac{e^{\lambda t}-e^{C(x)t}}{\lambda -C(x)} , $$
where $C(x) =\beta _0(N(x,0) +n_0\frac{x}{m})+ \mu +D_I$.\\
Note that $\widehat I_j, j=1,2$ satisfies
\begin{equation}\label{ichapeau}
\frac{\partial \widehat I_j}{\partial t} =\widehat I_j(-\mu+ \beta_0\widehat S_j)-D_I(\widehat I_j-\frac{x}{m}i_j(t)),
\end{equation}
which, using that $\widehat I_j +\widehat S_j =\widehat N$ by Proposition \ref{Nexpression}, can be rewritten as 
$$ \frac{\partial \widehat I_j}{\partial t}=- \beta_0\widehat I_j^2+\widehat I_j[\beta _0\widehat N-(\mu +D_I)] +D_I\frac{x}{m}i_j(t).$$
Thus $y$ satisfies 
\begin{eqnarray*}
\frac{\partial y}{\partial t}&=&   y(-\beta _0(\widehat I_1+\widehat I_2)+\beta _0 \widehat N(x,t)-( \mu +D_I)) +D_I\frac{x}{m}( i_2(t)-i_1(t))  \\
&=&  \alpha(x,t) y +D_I\frac{x}{m}( i_2(t)-i_1(t)),
\end{eqnarray*}
with
$$\forall (x,t) \in J_T, \,\,\,\alpha (x,t)= -\beta _0(\widehat I_1+\widehat I_2) + \beta _0\widehat N (x,t)-( \mu +D_I).$$
Since $0\leq \widehat I_j \leq \widehat N$ for $j=1,2$, we have that
 $$\forall (x,t) \in J_T, \,\,\,-\beta _0\widehat N (x,t)-( \mu +D_I)  \leq \alpha (x,t) \leq  \beta _0\widehat N (x,t)-( \mu +D_I)$$
so that it follows from (\ref{majhat}) that
 $$\forall (x,t) \in J_T, \,\,\,\vert \alpha (x,t) \vert \leq  \beta _0\widehat N (x,t) + \mu +D_I \leq  \beta _0(N(x,0) +n_0\frac{x}{m})+ \mu +D_I.$$
Hence
\begin{equation}\label{eqI1et2}
 \vert\frac{\partial y}{\partial t}   \vert \leq  \,\, C(x) \vert  y\vert+ A(x,t),
 \end{equation}
with
 $$C(x) =   \beta _0(N(x,0) +n_0\frac{x}{m})+ \mu +D_I \mbox{ and }A(x,t)=D_I\frac{x}{m}\vert i_2(t)-i_1(t)\vert . $$
Note that $y(x,0)=0$ for all $x \in J$.
 Hence we can use Theorem \ref{LG} and deduce that for all $(x,t) \in J_T$, 
  \begin{equation}\label{cony}
 \vert y(x,t)\vert \leq e^{C(x)t}\int_0^{t} e^{-C(x) s} A(x,s) ds 
  \end{equation}
  Let us choose  $\lambda > M=\max_{[0, \xmax]}C(x)$.
In view of the definition of the norm in $E$, we have that for all $s \in [0,T]$ and $x \in J$,
$$ 0 \leq A(x, s) \leq D_I\frac{x}{m}  \Vert i_2-i_1\Vert e^{\lambda s} $$
so that 
$$\int_0^{t} e^{-C(x)s} A(x,s) ds \leq D_I\frac{x}{m}  \Vert i_2-i_1\Vert \frac{e^{(\lambda -C(x))t}-1}{\lambda -C(x)}.$$
 Therefore it follows from (\ref{cony}) that
$$\vert y(x,t)\vert \leq D_I\frac{x}{m} \Vert i_2-i_1\Vert \frac{e^{\lambda t}-e^{C(x)t}}{\lambda -C(x)} \leq D_I\frac{x}{m} \Vert i_2-i_1 \Vert \frac{e^{\lambda t}}{\lambda -M}.$$
Hence
$$\vert \widehat{i}_2(t)-\widehat{i}_1(t)\vert \leq \int_0^{\xmax} \vert y(x,t)\vert  p(x)dx \leq  D_I \Vert i_2-i_1 \Vert \frac{e^{\lambda t}}{\lambda -M}  $$
 so that finally  
   $$ \Vert \widehat{i}_2-\widehat{i}_1\Vert \leq  \frac{D_I}{\lambda -M}\Vert i_2-i_1 \Vert .$$
 We choose $\lambda > M+D_I$ large enough so that $k =  \frac{D_I}{ \lambda -M} < 1$.
Then $\phi$ is a contraction in $E$. Therefore it admits a unique fixed point, i.e. a function $i:[0,T] \rightarrow \R_+$ 
such that $\widehat i=i$ on $[0,T]$. Since in view of  Lemma  \ref{nochapeau}, we have that $s(t) + i(t) = \widehat s(t) + \widehat i(t) = n_0$  on $[0,T]$, it follows that 
 $\widehat s = s$  on $[0,T]$. Thus the corresponding function pair $(\widehat S, \widehat I)= (S,I)$ is the unique solution to system $(C)$ on $[0,T]$.

\subsection{Existence of solutions to $(C)$ in the general case $D_I, D_S>0$} 
Define 
 \begin{equation}
 d= \min( D_I, D_S), \,\,\,\,\,\, D= \max( D_I, D_S).
 \label{Dd}
 \end{equation}
 We first establish some estimates on the solutions to system $(\widehat C)$ .
\begin{proposition}\label{Nchapeau}
Let $T>0$ be given and consider 2  smooth nonnegative functions $(s,i)$ defined on $[0,T]$. We assume  that there exists $R\geq 0$ so that
$$\forall t  \in [0,T],\,\,\, 0\leq s(t)+i(t) \leq n_0 + R $$ with $n_0=s(0)+i(0)$ .
Let  $(\widehat S,\widehat I)$ be the solution to system $(\widehat C)$  on $\R_+ \times [0,T]$ and let
$\widehat N = \widehat S + \widehat I$.
Then  
\begin{equation}
\forall (x,t) \in J_T, \,\,\, 
0 \leq \widehat N(x,t)\leq K(x) e^{-dt}+\frac{D}{d}(n_0 + R)\frac{x}{m},
\label{Nub}
\end{equation}
 with  $K(x)=N(x,0)-\frac{D}{d}(n_0 +R )\frac{x}{m}$. 
 In particular,
\begin{equation}
\forall (x,t) \in J_T, \,\,\, 0 \leq \widehat N(x,t)\leq N_+(x) ,
\label{majN}
\end{equation}
where
 \begin{equation}
\forall x \in J, \,\,\, 
N_+(x)= \frac{D}{d} (n_0 + R)\frac{x}{m}+ N(x,0)
\label{N+}
\end{equation}
and 
$$\forall t  \in [0,T], \,\,\, \widehat n(t)\leq \frac{D}{d} (n_0+R).$$
\end{proposition}
{\bf Proof}\\
Adding up the two equations in $(\widehat C)$ yields
\begin{eqnarray*}
\frac{\partial \widehat N}{\partial t} &=&-D_S(\widehat S-\frac{x}{m}s(t))-D_I(\widehat I-\frac{x}{m}i(t)) \\
&=&-D_S\widehat S -D_I\widehat I   +D_S\frac{x}{m}s(t)+D_I\frac{x}{m}i(t).
\end{eqnarray*}
Since the functions $\widehat S$, $\widehat I$, $ s(t)$, $i(t) $ are nonnegative and $d\leq D_I, D_S \leq D$,  it follows that
 $$\frac{\partial \widehat N}{\partial t}  \leq -d \widehat N(x,t) + D \frac{x}{m}(s(t)+i(t)) \leq -d \widehat N(x,t) + D (n_0 + R)\frac{x}{m}.$$
 Hence by standard ODE comparison theorem, 
 \begin{equation}\label{supNhat}
  \forall (x,t) \in J_T, \,\,\, 
0 \leq \widehat N(x,t)\leq  N^+(x,t),
\end{equation} where
$$
\left\{
\begin{array}{lcll}
\displaystyle{\frac{\partial N^+}{\partial t}}&=&  \displaystyle{-d N^+ + D(n_0 + R) \frac{x}{m} }   & x \in J, \, t \in [0,T] \\
\vspace{2mm}
 N^+(x,0)&=&N(x,0),& x \in J,
\end{array}
\right.
$$
 A straightforward computation yields
 $$\forall (x,t) \in J_T, \,\,\,N^+(x,t)= K(x) e^{-dt}+\frac{D}{d}(n_0+R)\frac{x}{m}$$
where
 $K(x)= N(x,0) -\frac{D}{d}(n_0 + R)\frac{x}{m} $, which yields (\ref{Nub}) by (\ref{supNhat}).
Note that
 $$N^+(x,t)= N(x,0)e^{-dt}+(1-e^{-dt})\frac{D}{d}(n_0 + R )\frac{x}{m} \leq N_+(x),$$
 with $N_+$ defined in (\ref{N+}) so that we have established (\ref{majN}).
 Consequently, using that $\frac{D}{d} \geq 1$ and $R\geq 0$,
$\widehat n$ satisfies
$$\forall t \in [0,T], \,\,\, 0 \leq \widehat n(t)\leq n_0 e^{-dt} + \frac{D}{d}(n_0+R) (1- e^{-dt} ) \leq\frac{D}{d}(n_0+R) $$

\noindent
Let  us  consider  the set 
\begin{eqnarray*}
E &=&  \lbrace (s,i) \in C^0([0,T],\R^2), \forall t \in [0,T],  s(t) \geq 0, i(t) \geq 0, \\
&&  s(0) = s_0, i(0)=i_0,  \Vert (s,i)(t)-(s_0,i_0)\Vert \leq R \rbrace
\end{eqnarray*}
for some $R\geq1$ to be defined later. Here, $C^0([0,T],\R^2)$ is equipped with the  norm $\Vert (f,g) \Vert =\displaystyle{\sup_{t\in [0,T]}(\vert f(t)\vert +\vert g(t)\vert})$. 
This set is a closed, convex subset of $ C^0([0,T],\R^2)$. The corresponding function $n:[0,T]\rightarrow \R_+$ is then defined by $n=s+i $ on $[0,T]$. 
\begin{proposition}\label{H_2}
We define $$\phi:\,\ E \longrightarrow C^0([0,T],\R^2) $$ 
                  $$       (s,i)          \longmapsto    (\widehat s,\widehat i), $$ 
Then for $T>0$ small enough, $ \phi (E)\subset E$ and  
$\phi $ is a contraction in $E$.                 
\end{proposition}
{\bf Proof of Proposition \ref{H_2}}

\noindent
We first show that $\phi(E) \subset E$.  Let us consider  $(s,i)\in E$ and show that  $ (\widehat s,\widehat i)= \phi( s,i ) \in  E$.
Note that by Lemma \ref{signeichapeau},  $\widehat{i}, \widehat{s}\geq 0$ on $[0,T]$. 
First we obtain an upper bound for  $ \vert \widehat{i}'(t)\vert $. The second equation in System $(\widehat c)$  reads
$$\widehat{i}'(t)=-(\mu + D_I)\widehat{i}(t)+\beta_0 \int_0^{\xmax}(\widehat{S}\widehat{I})(x,t) \, p(x)dx +D_Ii(t) $$
Thus 
\begin{equation}\label{iprime}
\vert \widehat{i}'(t)\vert \leq (\mu +D_I)\vert \widehat{i}(t)-i(0)\vert +\mu i(0) +\beta_0  \int_0^{\xmax}( \widehat{S}\widehat{I}) (x, t)p(x)dx  +D_I\vert i(t)-i(0)\vert 
\end{equation}
Since $0 \leq \widehat{S}\widehat{I} \leq \widehat{N}^2$, it follows from Proposition \ref{Nchapeau} that
$$0 \leq \int_0^{\xmax}(\widehat{S}\widehat{I})(x,t) \, p(x)dx  \leq \int_0^{\xmax}N_+^2(x) \, p(x)dx.$$ 
Let us denote 
$$ C_1= \int_0^{\xmax}x^2p(x)dx  \mbox{ and } C_2=\int_0^{\xmax}N^2(x,0)p(x)dx $$
so that since $R\geq 1$,
$$ \int_0^{\xmax}N_+^2(x) \, p(x)dx \leq   2(\frac{D}{d}\frac{(n_0+R)}{m})^2C_1+2 C_2 \leq Z_0 R^2 $$
for some constant  $Z_0>0$. Hence, using that  $ \vert i(t)-i(0)\vert \leq R \leq R^2$, we deduce from (\ref{iprime}) that
\begin{eqnarray*} 
\vert \widehat{i}'(t)\vert &\leq & (\mu + D_I) \vert\widehat{i}(t)-i(0)\vert+ \mu i(0)+ D_I\vert i(t)-i(0)\vert 
 + \beta_0 Z_0 R^2 \\
  &\leq & (\mu + D_I) \vert\widehat{i}(t)-i(0)\vert+Z_1 R^2
 \end{eqnarray*} 
 for some constant  $Z_1>0$.
Let us denote
$$f(t)= \widehat{i}(t)-i(0),\,\,  C=\mu + D_I,  $$
so that $f$ satisfies $f(0)=0$ and 
$$ \forall t \in [0,T], \,\,\, \vert f'(t)\vert \leq  C \vert f(t) \vert+  Z_1 R^2 .$$
Applying Theorem \ref{LG} to the function $f$, we obtain that
$$ \forall t \in [0,T], \,\,\, \vert f(t)\vert \leq  \,\,\,   Z_1 R^2 e^{C t}\int_0^t e^{-C u}du  .$$
Finally, we have that 
$$ \forall t \in [0,T], \,\,\, \vert f(t)\vert \leq  \,\,\,    Z_1 R^2 \frac{e^{Ct}-1}{C} \leq \frac{R}{2} $$
if $T>0$ is small enough. 
Similarly, we can show that for if $T>0$ is small enough,
$$ \forall t \in [0,T], \,\,\, \vert \widehat{s}(t)-s_0 \vert \leq  \,\,\, \frac{R}{2}$$
so that
$$ \forall t \in [0,T], \,\,\, \Vert (\widehat{s}, \widehat{i})-(s_0,i_0) \Vert \leq  \,\,\, R$$
which shows that $(\widehat{s}, \widehat{i}) \in E$.

b) {\bf $\phi$ is a contraction.}

Next, let us consider $(s_1,i_1) \in E$ (resp. $(s_2, i_2) \in E$) and denote by  $(\widehat S_1,\widehat I_1)$ (resp. $(\widehat S_2,\widehat I_2) )$ the corresponding solutions of system $(\widehat C)$ and define accordingly $\widehat N_j= \widehat S_j +  \widehat I_j$ for $j=1,2$.
Let us define the functions $y: [0, \xmax] \times [0,T]  \rightarrow \R$ and  $z: [0, \xmax] \times [0,T]  \rightarrow \R$ by
$$\forall x \in [0, \xmax], \forall t \in  [0,T],\,\,\,  y(x,t) = (\widehat I_2- \widehat I_1)(x,t), \,\,\,   z(x,t) = (\widehat N_2- \widehat N_1)(x,t).$$
We establish the following estimates on $y$ and $z$.

\noindent
In a first step, we prove that  for all $(x,t) \in \R_+\times [0,T]$, 
\begin{equation}
 \vert\frac{\partial y}{\partial t} \vert  \, \leq  \, C_1(x) (\vert  y\vert+  \vert  z\vert )+  D_I\frac{x}{m}\vert i_2-i_1\vert (t) 
 \label{diffbigI2}
\end{equation}
with $ C_1(x)=2\beta _0 N_+(x)+\mu +D_I $.

\noindent
Note that $\widehat I_j$ ($j=1,2$) satisfies
$$\frac{\partial \widehat I_j }{\partial t} =\widehat I_j(-\mu+  \beta_0 \widehat S_j) -D_I(\widehat I_j-\frac{x}{m}i_j(t)).$$
Using that $\widehat S_j = \widehat N_j -\widehat I_j$, we obtain that
$$\frac{\partial \widehat I_j}{\partial t} =\widehat I_j(-\mu+ \beta_0 \widehat N_j) -  \beta_0 (\widehat I_j)^2 -D_I (\widehat I_j -\frac{x}{m}i_j(t)).$$
Thus, if we define $\Delta$ by 
 $$\forall (x,t) \in  \R_+\times [0,T],\,\,\, \Delta(x,t) = \beta_0 (\widehat N_2 \widehat I_2
-  \widehat N_1 \widehat I_1) (x,t)  -  \beta_0  ((\widehat I_2)^2
- (\widehat I_1)^2 ) (x,t) $$
then $y$ satisfies on $\R_+ \times [0,T]$ 
\begin{equation}\label{icha2}
\frac{\partial y }{\partial t} = 
 \Delta  -(\mu +D_I) y +D_I\frac{x}{m}(i_2(t)-i_1(t)),
\end{equation}
We omit $(x,t)$ and rewrite 
\begin{eqnarray*}
\Delta &= &\beta_0 \widehat N_2 (\widehat I_2-\widehat I_1) + \beta_0 \widehat I_1(\widehat N_2 -\widehat N_1)-  \beta_0   (\widehat I_2
+\widehat I_1 ) (\widehat I_2
- \widehat I_1 ) \\
 &= &\beta_0 ( \widehat S_2 - \widehat I_1) y + \beta_0 \widehat I_1 z 
 \end{eqnarray*}
which we substitute in (\ref{icha2}) and obtain that 
$$\frac{\partial y}{\partial t} =( \beta_0 ( \widehat S_2  - \widehat I_1)-(\mu +D_I) )y  + \beta_0 \widehat I_1z  +D_I\frac{x}{m}(i_2-i_1)(t).$$
Since $0\leq \widehat I_1 \leq \widehat N_1$ and $0\leq \widehat S_2 \leq \widehat N_2$, it follows that
$$\vert \frac{\partial y}{\partial t} \vert \leq (\beta_0 (\widehat N_2 +\widehat N_1) +\mu +D_I ) \vert  y \vert  + \beta_0 \widehat N_1\vert z \vert + 
D_I\frac{x}{m}\vert i_2-i_1\vert (t) $$
which yields (\ref{diffbigI2}) in view of Proposition \ref{Nchapeau}, since we have that 
 $$\forall x \in [0,x_{max}], \,\,\, \forall t \in [0,T], \,\,\, 0 \leq  (\widehat N_2 +\widehat N_1) (x,t)\leq  2 N_+(x).$$
 
 \noindent
In a second step, we prove a similar differential inequality on $z=\widehat N_2 - \widehat N_1$ and show
that
 for all $(x,t) \in \R_+\times [0,T]$, 
\begin{equation}
 \vert\frac{\partial z}{\partial t} \vert  \, \leq  \, D (\vert  y\vert+  \vert  z\vert )+  D\frac{x}{m}(\vert i_2-i_1\vert (t)+  \vert n_2-n_1\vert (t))\label{diffbigN2}
\end{equation}
with $D= \max(D_I,D_S)$.

\noindent
Note that adding up the two equations in system $(\widehat C)$ shows that $\widehat N_j$ ($j=1,2$) satisfies
\begin{eqnarray*}
\frac{\partial \widehat N_j }{\partial t} &=-&D_S(\widehat S_j - \frac{x}{m}s_j(t) )-D_I(\widehat I_j(x,t)  - \frac{x}{m}i_j(t)) \\
 &=-&D_S(\widehat N_j - \frac{x}{m}n_j(t) )-(D_I - D_S)(\widehat I_j(x,t)  - \frac{x}{m}i_j(t)).
 \end{eqnarray*}
By substraction, it follows that
$$ \frac{\partial z}{\partial t}=-D_S z
-(D_I-D_S) y+ D_S\frac{x}{m}(n_2-n_1)(t) +(D_I-D_S) \frac{x}{m}(i_2-i_1)(t)$$
 which yields (\ref{diffbigN2}) using that $0 \leq D_S, \vert D_I-D_S\vert \leq D$.

\noindent
Finally, if we denote 
$$C(x)= C_1(x) + D \mbox{ and }A(x,t)=2 D\frac{x}{m}(\vert i_2-i_1\vert(t) +  \vert n_2-n_1 \vert(t) ) ,$$
 we have established that 
there exists $C>0$ such that  for all $(x,t) \in [0, \xmax] \times [0,T]$, 
\begin{equation}
 \vert\frac{\partial y}{\partial t} \vert  +  \vert\frac{\partial z}{\partial t} \vert  \, \leq  \, C(x) (\vert  y\vert+  \vert  z\vert )+  A(x,t) \label{both2}
\end{equation}
Let us define the function $F$ for $ (x,t) \in [0, \xmax] \times [0,T]$ by
$$F(x,t) = \int_0^t \vert  \frac{\partial y}{\partial t}(x,\tau)\vert  +\vert  \frac{\partial z}{\partial t}(x,\tau)\vert d\tau$$ 
so that $A,F\geq 0$ and $F(x,0) =0$. Since $y(.,0)=z(.,0)=0$, $ \vert  y(x,t)\vert  +  \vert  z(x,t)\vert  \leq F(x,t)$ so that the inequality (\ref{both2}) implies 
\begin{equation}\label{F}
\forall (x,t) \in [0, \xmax] \times [0,T],\,\,\,  \frac{\partial F}{\partial t}(x,t) \leq C(x) F(x,t) + A(x,t).
\end{equation}
By Gronwall's lemma, this implies that
$$\forall (x,t) \in [0, \xmax]\times [0,T],\,\,\, \vert  y(x,t)\vert +  \vert  z(x,t)\vert \leq F(x,t)\leq \int_0^{t} e^{C(x) (t-s)} A(x,s)ds. $$
Since for all $s \in [0,T]$,
$$ \,\,\,\vert i_2(s)-i_1(s)\vert + \vert n_2(s)-n_1(s)\vert \leq 2 \Vert (s_2-s_1, i_2-i_1) \Vert ,$$ it follows that since $C(x)\neq 0$,
$$\forall (x,t) \in  [0, \xmax] \times [0,T],\,\,\, F(x,t) \leq 4D \frac{x}{m}  \Vert (s_2-s_1, i_2-i_1) \Vert   \frac{e^{C(x) t}- 1}{C(x) }$$
Note that by definition of $\widehat i$ and $\widehat s$, we have  that for all $t \in [0,T]$, 
$$\vert \widehat{i}_2(t)-\widehat{i}_1(t)\vert= \vert \int_0^{\xmax}  y(x,t) p(x) dx \vert \leq \int _0 ^{\xmax}\vert y(x,t) \vert p(x)dx $$ and
$$\vert \widehat{s}_2(t)-\widehat{s}_1(t)\vert= \vert \int_0^{\xmax}  (z-y)(x,t) p(x) dx \vert \leq \int _0 ^{\xmax} (\vert y(x,t) \vert +  \vert z(x,t) \vert) p(x)dx. $$
Thus multiplying the above inequality  by $p(x)$ and integrating on $[0,\xmax]$, we obtain that for all $t \in [0,T]$, 
$$\vert \widehat{s}_1(t)-\widehat{s}_2(t)\vert + \vert \widehat i_1(t)-\widehat i_2(t)\vert
\leq 8 D \Vert (s_2-s_1, i_2-i_1) \Vert    \int_0^{\xmax} \frac{e ^{C(x)t}-1}{C(x)} \frac{x}{m} p(x)dx . $$
For any fixed $R\geq 1 $, let $ M=\max_{[0, \xmax]}C(x)$ so that  $0<D \leq C(x)\leq M$ for $x \in [0,x_{max}]$.
Thus 
$$\forall (x,t) \in [0, \xmax]\times [0,T],\,\,\,   0<\frac{e ^{C(x)t}-1}{C(x)}\leq \frac{e ^{M T} - 1 }{D}$$ 
 and finally 
 $$\forall t \in  [0,T],\,\,\,  \Vert (\widehat s _2 -\widehat s_1, \widehat i _2 -\widehat i_1  )\Vert \leq  8 ( e ^{M T} - 1 ) \Vert (s_2-s_1, i_2-i_1)\Vert  $$
 so that $\phi$ is a contraction if we choose $T >0$ small enough (depending on $R$) in order to guarantee that $k = 4 (e ^{M T}-1)<1$.

\section{Existence and linear stability of $(DFE)$}
In this section, we consider equilibrium solutions to system (C).
Hence we look for nonnegative functions $(S^*(x), I^*(x))$ which satisfy  system $(C^*)$ on $\R_+$
$$(C^*)
\left\{
\begin{array}{lcll}
I(x)^*(\mu-\beta_0 S(x)^*)&=&D_S(S(x)^*-\frac{x}{m}s^*)\\
I(x)^*(-\mu+\beta_0 S(x)^*)&=&D_I(I(x)^*-\frac{x}{m}i^*).
\end{array}
\right.
$$
with
\begin{equation}
s^*=\int_0^{\xmax }S^*(x)p(x)dx \mbox{ and } i^*=\int_0^{\xmax }I^*(x)p(x)dx.
\label{eq-closure}
\end{equation}
The disease-free and endemic equilibria are defined as follows.
\begin{definition} For a given $n_0>0$,  any equilibrium solution such that $<N^*(x)>= n_0$  is of one of the two following types,
\begin{itemize}
\item The disease-free equilibrium (DFE), given by
  $$\forall x \geq 0,\,\,\, I^*(x)=0, \,\,\, S^*(x) = n_0 \frac{x}{m},\,\,\, $$
\item An endemic equilibrium (EE),  which is a nonnegative solution  $(S^*(x),I^*(x))$ such that there exists $x_0>0$ with
$I^*(x_0) >0$.
\end{itemize}
\end{definition}
Let us remark there always exists a $(DFE)$.\\
\subsection{Necessary condition of linear stability}
Let $(S^*(x), I^*(x))$ be an equilibrium solution to system $(C)$, with $ s^*=<S^*>$,  $ i^*=<I^*>$ and $n_0= s^* +i^*$. The linearized system around $(S^*(x), I^*(x))$ is given for all $x>0$ by
$$(L)
\left\{
\begin{array}{lcr}
\frac{\partial f}{\partial t} &=&-\beta I^*(x) f  + (\mu -\beta S^*) g- D_S(f - \frac{x}{m}\int_0^{x_{max}} f(x,t) p(x)dx)\\
\frac{\partial g }{\partial t} &=&\beta I^*(x) f+(-\mu+\beta S^*)g-D_I(g - \frac{x}{m}\int_0^{x_{max}} g(x,t) p(x)dx),
\end{array}
\right.
$$
 with given initial data $(f(x,0), g(x,0))= (f_0(x), g_0(x))$, 
   which are 2  functions from $\R_+$ to $\R$ such that  
 $$\int_0^{x_{max}} \vert f_0(x) \vert 
p(x) dx < \infty, \,\,\,  \int_0^{x_{max}} \vert g_0(x) \vert 
p(x) dx < \infty.$$
We make the assumption that $f_0, g_0: \R_+ \rightarrow \R$ are $C^1$ and satisfy
\begin{equation}
 \langle f_0 \rangle +  \langle  g_0 \rangle = \int_0^{\xmax}(f_0(x)+g_0(x)) \,  p(x) \,dx=0.
\label{ncstab}
\end{equation}
We define linear stability of $(L)$ as follows:\\
\begin{definition}
$(S^*, I^*)$ is linearly stable if and only if
 $\forall x>0$, 
$$\lim_{t\longrightarrow \infty } f(x,t)=\lim_{t\longrightarrow \infty } g(x,t)=0,$$ where $(f,g)$ is the solution of $(L)$ with  initial data $ (f_0, g_0)$.
\end{definition}

Indeed, for any solution $(f(x,t), g(x,t))$ to the linearized system $(L)$, let us define the functions $ h$ and $H$ by 
$$ \forall t\geq 0,\,\,\, h(.,t) = f(.,t) + g(.,t) \mbox{ and } H(t) =   \langle h(.,t) \rangle=\int_0^{\infty} h(x,t) \,  p(x) \,dx .$$ 
Note that $h$ satisfies
\begin{equation}\label{H}
 \frac{\partial h}{\partial  t}= -D_S(f - x/m  \langle f(.,t) \rangle) -D_I(g - x/m  \langle  g(.,t) \rangle) . 
\end{equation}
After multiplying (\ref{H}) by $p(x)$ and integrating on $\R_+$, we obtain that for any $t > 0$, $H'(t) = 0$.
Thus $H(t)=H(0)$ for all $t>0$ so that stability imposes $\displaystyle{\lim_{t\rightarrow \infty} H(t)}=H(0)=0$ which is condition (\ref{ncstab}).
 We also established that any solution $(f,g)$ to $(L)$ with $(f_0,g_0)$ satisfying (\ref{ncstab}) has the property that 
 \begin{equation}\label{cons}
  \forall t\geq 0,\,\,\, H(t) =  \langle f(.,t) \rangle +   \langle g(.,t) \rangle=0 .
  \end{equation}

\subsection{Linear stability of $(DFE)$ in general case}

\begin{theorem}\label{deh2}
Let us assume that $\frac{\beta _0x_{max}n_0}{m}<\mu$. 
Let $(f,g)$ be the solution of $(L)$ with  initial data $ (f_0, g_0)$.
Then  $\int _0^{x_{max} }\vert g(x,t)\vert p(x)dx<\infty $ for all $t>0$ and 
$$\forall x \in [0, \xmax], \,\,\, 
\lim_{t\longrightarrow \infty } g(x,t)=0 \mbox{ and } \lim_{t\longrightarrow \infty } f(x,t)=0 $$
which proves that the disease free equilibrium $(DFE)$ is linearly stable.  
\end{theorem}

{\bf proof of theorem \ref{deh2}}
a) We first consider the case where $g_0 \geq 0$ on $[0, \xmax] $ with $ \langle  g_0 \rangle >0$.
The function g satisfies
\begin{equation}\label{g+}
 \frac{\partial g}{\partial t}  (x,t)= g(-\mu+\frac{\beta _0xn_0}{m}-D_I)+D_I\frac{x}{m}\int_0^{x_{max}} g(x,t) p(x)dx 
 \end{equation}
We first prove for all $ x \in [0, \xmax]$ and $t> 0$, $g(x,t)>0$.\\
 
Let us define the function $G$  for $ t > 0$ by  $ G(t)= \int_0^{x_{max}} g(x,t)p(x)dx$ and let us denote  $A = \lbrace t\geq 0,\forall s \in [0,t],\ G(s)>0\rbrace $.
Since $G(0) = \langle  g_0 \rangle >0$ by assumption, $0 \in A$ so that A is not empty. 
Let us suppose that A is bounded by above and let us define $T^* = \sup(A)$ 
so  that $G(T^*) = 0$.\\
We define  for all $ x \in [0, \xmax]$ the function $r$ by
$r(x)=(-\mu+\frac{\beta _0xn_0}{m}-D_I) $.
Note that for all  $ x \in [0, \xmax]$,
$$\frac{\partial (e^{-r(x)t }g(x,t))}{\partial t}=e^{-r(x) t }[ -g (x,t)r(x,t)+ \frac{\partial g(x,t)}{\partial t}] = \frac{\partial( e^{-r(x)t}g(x,t))}{\partial t}=e^{-r(x) t}D_I\frac{x}{m}G(t) $$
so that or all  $t\in [0,T^*) $, $\frac{\partial e^{-r(x)t}g(x,t)}{\partial t}> 0$.
Since $g(x,0)\geq 0$, it follows that for all  $ x \in [0, \xmax]$ and  $ t \in [0,T*)$, $ e^{-r(x)t}g(x,t)> 0$ hence $g(x,t)>0$ and that 
 $$\forall x \in [0, \xmax], \,\,\,  \forall t \in [T^*/2,T^*), \,\,\,  e^{-r(x)t}g(x,t) \geq e^{-r(x) T*/2}g(x, T^*/2) >0.$$
 Thus $g(x,T*) > 0$ for all $ x \in [0, \xmax]$. This contradicts $G(T^*) = 0$ and proves that  $T^* =\infty$. Thus $G(t)>0$ for all $t \geq 0$ which in view of (\ref{g+}) 
 implies that $g(x,t)>0$ for all $ x \in [0, \xmax]$ and $t> 0$.

It follows that  $0\leq g(x,t) \leq g_+(x,t)$ where $g_+$ is the solution of
$$\frac{\partial g_+ }{\partial t} (x,t)= g_+(-\mu+\frac{\beta _0x_{max} n_0}{m}-D_I)+D_I\frac{x}{m}\int_0^{x_{max}} g_+ p(x)dx.$$
Let us multiply by $p(x)$ and integrate on $ [0, x_{max}]$ and define the function $G_+$ for $ t > 0$ by $ G_+(t)= \int_0^{x_{max}} g_+(x,t)p(x)dx$. Then 
$$G_+'(t)=(-\mu+\frac{\beta _0x_{max}n_0}{m}-D_I)G_+(t)+D_IG_+(t) $$
so that  $G_+'(t)=(-\mu+\frac{\beta _0x_{max}n_0}{m})G_+(t)$. Hence, $G_+(t)=G(0)e^{(-\mu+\frac{\beta _0x_{max}n_0}{m})t}$. Since for all $t>0$,
$$0 \leq G(t)  \leq G_+(t) < \infty,$$
this proves that $\int _0^{x_{max} }\vert g(x,t)\vert p(x)dx<\infty $.

Next we compute the supersolution $g+$
 for all $x \in [0, \xmax]$ and $t >0$ and obtain 
$$g_+(x,t)=g_0(x)e^{(-\mu+\frac{\beta _0x_{max}n_0}{m}-D_I)t}+\frac{ x_{max}n_0}{m}G(0)e^{(-\mu+\frac{\beta _0x_{max}n_0}{m})t}. $$
This function converges to 0 when $-\mu+\frac{\beta _0x_{max}n_0}{m}< 0$.
This implies that 
$$\forall x \in [0, \xmax], \,\,\, 
\lim_{t\longrightarrow \infty } g(x,t)=0. $$

b) In the case when $g_0$ might change sign, we write $g_0$ as $g_0(x)=g^+_0(x)-g^-_0(x)$, with $g^+_0$ and $g^-_0$ 2 nonnegative functions. 

By linearity of (\ref{g+}), we have that $g$ can be written as $g(x,t)=g^+(x,t)-g^-(x,t)$ where $g^\pm(x,t)$ is the corresponding solution of (\ref{g+}) with initial data $g_0^{/pm}$.
It follows from a) that  $g_{\pm}(x,t) $ converge to $0$ as $t \rightarrow \infty$ for all 
$x \in [0, \xmax]$,  which proves the result for $g$.

c) Note that in view of (\ref{H}) and (\ref{cons}), the function $h$  satisfies
\begin{equation}\label{H2}
 \frac{\partial h}{\partial  t}= -D_S h  + (D_S-D_I) (g - x/m  G(t)) . 
\end{equation}
A straightforward integration shows that for all $x \in [0, \xmax]$, 
$\lim_{t\longrightarrow \infty } h(x,t)=0$, which implies the same result for $f=h-g$.
Thus the (DFE) equilibrium is stable when $\frac{\beta _0x_{max}n_0}{m}<\mu$.

\section{Asymptotic behaviour }
In this section, we prove that when $n_0= \langle N(x,0) \rangle >0$ is small enough, the solution $(S,I)$ of $(C)$ converges to the $(DFE)$. (Note that we establish in \cite{h2equilibrium} that there is no endemic equilibrium if $n_0$ is small enough.)

\noindent
Let us first notice that there exists $C\geq 1$ such that 
\begin{equation}
\forall x \in J, \,\,\, 0 \leq N(x,0) \leq C n_0 \frac{x}{m} .
\label{C0}
\end{equation}
We now recall  the following lemma established in \cite{ipel1}, which  shows that it is sufficient to obtain the limiting behavior of $I(x,t)$ as $t \rightarrow \infty$.

\noindent
{\bf Lemma 5.1}  \cite{ipel1}.

\noindent
Assume that the initial data $\left(S_0,I_0\right)$  satisfy $(ID1)$ and $(ID2)$ and (\ref{C0}). Then the following properties hold.
\begin{enumerate}
\item Define 
$\tilde C= \max(C,\frac{D}{d}) \geq 1 $. Then for all $x \in J$ and for all $t \geq 0$,
\begin{equation}
0 \leq N(x,t)\leq \tilde C  n_0\frac{x}{m}.
\label{Nmaj}
\end{equation}
\item For any fixed $x \in J$, we have that
$$
\lim_{t\rightarrow \infty } (I(x,t) - \frac{x}{m} i(t))=0   \,\,\,\, \Rightarrow  \lim_{t\rightarrow \infty }( N(x,t) -n_0\frac{x}{m}) = 0$$
$$ \,\,\,\, \Rightarrow  \lim_{t\rightarrow \infty }( S(x,t) -\frac{x}{m} s(t)) = 0$$
\end{enumerate}
Next we establish the following result.
\begin{theorem}\label{cvel}
Assume that $n_0<\frac{1}{\tilde C}  \frac{m}{\xmax}  \frac{\mu}{\beta_0}$, where $d$, $D$ are defined in (\ref{Dd}). Then the solution to system $(C)$ converges to the $(DFE)$. Precisely,
$$\forall x \in J,\,\,\, \lim_{t\longrightarrow \infty} I(x,t)=0
\mbox{ and }
 \lim_{t\longrightarrow \infty}S(x,t)=n_0\frac{x}{m}.$$
\end{theorem}
Using (\ref{Nub})  and (\ref{C0}) in the case
where $(\widehat S,\widehat I)= (S,I)$, $\widehat N =N$  and $R=0$, we obtain the following estimates of the solution $(S,I)$ and of $N= S +I$,
\begin{equation}
\forall (x,t) \in J \times \R_+, \,\,\, 0 \leq N(x,t)\leq \max(C,\frac{D}{d})  n_0\frac{x}{m}.
\label{majN2}
\end{equation}
Since $I$ satisfies 
\begin{equation}\label{I}
\dfrac{\partial  I}{\partial t}=I(-\mu+\beta_0  S)-D_I( I-\frac{x}{m}i(t)) \mbox{ on } J \times \R_+
\end{equation}
it follows that 
$i(t)=<I(x,t)>$ satisfies
$$\forall t >0, \,\,\, i'(t)=\int_0^{x_{max}}I(x,t)(-\mu+\beta_0 S(x,t)) p(x)dx$$
where by (\ref{majN2}) 
$$ \forall x \in J, \,\,\, -\mu+\beta_0 S(x,t)\leq -\mu+\beta_0 {\tilde C} n_0 \frac{x_{max}}{m}.$$
Thus $0 \leq i(t) \leq i_+(t)$ for all $t>0$, where
$$\frac{\partial  i^+}{\partial t}= i^+(t)(-\mu+ \beta_0 \tilde C n_0  \frac{x_{max}}{m}), \,\,\, 
i^+(0)=i_0.$$
Since $i^+(t)=i_0e^{(\beta_0 {\tilde C} n_0 \frac{x_{max}}{m}-\mu)t}$, it follows that 
$\lim_{t\rightarrow \infty}i(t)=0$  if $\tilde C\beta_0 n_0 \frac{x_{max}}{m}-\mu<0$.
Next it follows from (\ref{I}) that  
$ 0 \leq I \leq I^+ $ on $ J \times \R_+$, where 
$I^+$ satisfies 
$$\frac{\partial  I^+}{\partial t}=I^+(-\mu+\beta_0 \tilde C n_0 \frac{x_{max}}{m} )-D_I( I^+-\frac{x}{m}i^+(t)),$$ 
with $I^+(..0) = I(.,0) $ and $i^+(t)$ given above. 
A straightforward computation yields that for all $(x,t) \in J \times \R_+$,
$$I^+(x,t)= (I(x,0)-i_0\frac{x}{m})e^{(\beta_0 \tilde C n_0 \frac{x_{max}}{m}-\mu-D_I)t}+ i_0\frac{x}{m}e^{(\beta_0 \tilde C n_0 \frac{x_{max}}{m}-\mu )t}$$
so that 
$\lim_{t\rightarrow \infty}I(x,t)  =0$  if $\tilde C\beta_0 n_0 \frac{x_{max}}{m}-\mu<0$.
 In view of the above Lemma $5.1$, this implies that
 $$\lim_{t\longrightarrow \infty}S(x,t)=\lim_{t\longrightarrow +\infty}N(x,t)=n_0\frac{x}{m},$$
 which completes the proof of Theorem \ref{cvel}.

\end{document}